\documentclass[aps,prb,twocolumn,groupedaddress]{revtex4}
\usepackage{graphicx}
\usepackage{dcolumn}
\usepackage{bm}
\begin{document}

\title{Would Bohr be born if Bohm were born before Born?}

\author{Hrvoje Nikoli\'c}
\affiliation{Theoretical Physics Division, Rudjer Bo\v{s}kovi\'{c}
Institute,
P.O.B. 180, HR-10002 Zagreb, Croatia.}
\email{hrvoje@thphys.irb.hr}

\date{\today}

\begin{abstract}
I discuss a hypothetical historical context in which a Bohm-like
deterministic interpretation of the Schr\"odinger equation could have been
proposed before the Born probabilistic interpretation and argue that in such
a context the Copenhagen (Bohr) interpretation would probably have never
achieved great popularity among physicists.


\vspace*{0.6cm}

\begin{quote}                            
\hspace*{5cm} {\it Is this the real life \\
\hspace*{5cm} Is this just fantasy \\
\hspace*{5cm} Caught in a landslide \\
\hspace*{5cm} No escape from reality} \\
\\
\hspace*{5cm} Freddie Mercury, ``Boh(e)mian Rhapsody"
\end{quote}

\end{abstract}

\maketitle

\section{Introduction}

The Copenhagen interpretation of quantum mechanics (QM) was the first
interpretation of QM that achieved a significant recognition among
physicists. It was proposed very early by the fathers of QM, especially
Bohr and Heisenberg. Later, many other interpretations of QM were
proposed, such as statistical ensemble interpretation, Bohm (pilot wave)
interpretation, Nelson (stochastic dynamics) interpretation,
Ghirardi-Rimini-Weber (spontaneous collapse) interpretation, quantum logic
interpretation, information theoretic interpretation, consistent histories
interpretation, many-world (relative state) interpretation, relational
interpretation, etc. All these interpretations seem to be consistent with
experiments, as well as with the minimal pragmatic ``shut-up-and-calculate
interpretation". Nevertheless, apart from the minimal pragmatic
interpretation, the Copenhagen interpretation still seems to be the
dominating one. Is it because this interpretation is the simplest, the most
viable, and the most natural one? Or is it just because of the inertia of
pragmatic physicists who do not want to waste much time on (for them)
irrelevant interpretational issues, so that it is the simplest for them to
(uncritically) accept the interpretation to which they were exposed first? I
believe that the second answer is closer to the truth. To provide an argument
for that, in this essay I argue that if some historical circumstances
had been only slightly different, then it would have been very likely
that the
Bohm deterministic interpretation would have been proposed and 
accepted first,
and consequently, that this interpretation would have been 
dominating even today.\cite{foot0}
(In fact, if the many-world interpretation taken literally is correct, then
such an alternative history of QM is not hypothetical at all. Instead, it is
explicitly realized in many branches of the whole multi-universe containing
a huge number of parallel universes.) 
For the sake of easier reading, in the next section I no longer use the
conditional, but present an alternative hypothetical history of QM as if it
really happened, trying to argue that such an alternative history was
actually quite natural.\cite{foot1}
Although a prior knowledge on the Bohm deterministic interpretation
is not required here, for readers unfamiliar with this interpretation
I suggest to
read also the original paper \cite{bohm} or a recent pedagogic review
\cite{tumul}.

\section{An alternative history of quantum mechanics}

When Schr\"odinger discovered his wave equation, the task was to find an
interpretation of it. The most obvious interpretation -- that electrons are
simply waves -- was not consistent because it was known that electrons
behave as pointlike particles in many experiments. Still, it was known that
electrons also obey some wave properties. What was the most natural
interpretation of that? Of course, the notion of ``naturalness" is highly
subjective and strongly depends on personal knowledge, prejudices, and
current paradigms. At that time, classical deterministic physics was well
understood and accepted, so it was the most natural to try first with an
interpretation that maximally resembles the known principles of classical
mechanics. In particular, classical mechanics contains only real quantities,
so it was very strange that the Schr\"odinger equation describes a complex
wave. Consequently, it was natural to rewrite the Schr\"odinger equation in
terms of real quantities only. The simplest way to do this was to write the
complex wave function $\psi$ in the polar form $\psi=R e^{i\phi}$ and then
to write the complex Schr\"odinger equation as a set of two (coupled) real
equations for $R({\bf x},t)$ and $\phi({\bf x},t)$. However, such a simple
mathematical manipulation did not immediately reveal the physical
interpretation of $R$ and $\phi$. Fortunately, a physical interpretation was
revealed very soon, after an additional mathematical transformation 
\begin{equation}
\phi({\bf x},t)=\frac{S({\bf x},t)}{\hbar} ,
\end{equation}     
where $S$ is some new function. The Schr\"odinger equation
for $\psi$ rewritten in terms of $R$ and $S$ turns out to look remarkably
similar to something very familiar from classical mechanics. One equation
looks similar to the classical Hamilton-Jacobi equation for the function
$S({\bf x},t)$, differing from it only by a transformation
\begin{equation}
V({\bf x},t) \rightarrow V({\bf x},t) + Q({\bf x},t) ,
\end{equation}
where $V$ is the classical potential and 
\begin{equation}\label{Q}
Q \equiv -\frac{\hbar^2}{2m} \frac{\nabla^2 R}{R} .
\end{equation}
The other equation turns out to look exactly like the continuity equation 
\begin{equation}\label{cont}
\frac{\partial\rho}{\partial t}+\nabla(\rho {\bf v})=0
\end{equation}
for the density $\rho \equiv R^2$, with the Hamilton-Jacobi velocity
\begin{equation}\label{v}
{\bf v}=\frac{\nabla S}{m} .
\end{equation}
Thus, at that moment, the most natural interpretation of the phase of the
wave function seemed to be a quantum version of the Hamilton-Jacobi 
function that determines the velocity of a pointlike particle. But what was
$\rho$? Since one of the equations looks just like the continuity equation,
at the beginning it was proposed that $\rho$ was the density of particles.
That meant that the Schr\"odinger equation described a fluid consisting of a
huge number of particles. The forces on these particles depended not only on
the classical potential $V$, but also on the density $\rho$ through the
quantum potential (\ref{Q}) in which $R=\sqrt{\rho}$.\cite{foot2} 
    
Although the interpretation above seemed appealing theoretically, it was
very soon realized that it was not consistent with experiments. It could not
explain why, in experiments, only one localized particle at a single
position was often observed. Thus, $\rho$ could not be the density of a fluid.
It seemed that $\rho$ (or $R$) must be an independent continuous field,
qualitatively similar to an electromagnetic or a gravitational field, that,
similarly to an electromagnetic or a gravitational field, influences the
motion of a particle. But why does $\rho$ satisfy the continuity equation, 
what
is the meaning of this? They could not answer this question, but they were
able to identify a physical consequence of the continuity equation. To see
this, assume that one studies a statistical ensemble of particles with the
probability distribution of particle positions equal to some function
$p({\bf x},t)$. Assume also that, for some reason, the initial distribution
at $t=0$ coincides with the function $\rho$ at $t=0$. Then the continuity
equation implies that 
\begin{equation}\label{p=rho}
p({\bf x},t)=\rho({\bf x},t) 
\end{equation}
at {\em any} $t$. But why should these two functions coincide initially?
Although nobody was able to present an absolutely convincing 
explanation, at least
some heuristic arguments were found, based on statistical 
arguments.\cite{foot3}
This suggested that, in typical experiments, 
$\rho$ could be equal to the measured probability density of particle
positions. Indeed, it turned out that such a prediction agrees with
experiments. Since this prediction was derived from the natural assumption
that each particle has the velocity determined by (\ref{v}), it was
concluded that experiments confirm (\ref{v}). 
Thus, this interpretation
became widely accepted and received the status of an ``orthodox" 
interpretation.\cite{foot4}

However, not everybody was satisfied with this interpretation. In
particular, Born objected that there was no direct experimental evidence for
the particle velocities as given by (\ref{v}), so this assumption was
questioned by him. As an alternative, he proposed a different interpretation. In
his interpretation, the equality (\ref{p=rho}) was a fundamental postulate.
Thus, he avoided a need for particle velocities as given by (\ref{v}).
However, his interpretation has not been widely accepted among physicists.
The arguments against the Born interpretation were the following: First, this
{\it ad hoc} postulate could not explain {\em why} the probability density
was given by $\rho$. Second, a theory in which the probabilistic
interpretation was one of the fundamental postulates was completely against
all current knowledge about fundamental laws of physics. The classical
deterministic laws were well established, so it was more natural to accept a
deterministic interpretation of QM that differs from classical
mechanics 
less radically.
Third, it was observed that if one used the arguments of Born to argue that
QM is to be interpreted probabilistically, then one could use analogous
arguments to argue that even classical mechanics should be interpreted
probabilistically,\cite{foot5}
which seemed absurd. 

Although the Born purely probabilistic interpretation was not considered
very appealing, mainly owing to the overwhelming mechanistic view of physics
of that time, it was appreciated by some positivists that      
such an interpretation should not be excluded. The Born interpretation was
quite radical, but still acceptable as a possible alternative.
Indeed, his interpretation seemed to fit well with a
mathematically more abstract formulation of QM
(which started with the Heisenberg matrix formulation of QM
proposed even before the Schr\"odinger equation, and was further
developed by Dirac who formulated the transformation theory
and von Neumann who developed the Hilbert-space formulation),
in which Eq. (\ref{v}) did not seem very natural. 
However, one version of the Born interpretation was much more radical, in
fact too radical to be taken seriously. This new interpretation was
suggested by Bohr. In fact, Bohr was already known in the physics community 
for proposing the famous Bohr model of the hydrogen atom, in which electrons
move circularly at discrete distances from the nucleus. Now a much better
model of the hydrogen atom (the one based on the Schr\"odinger equation and
particle trajectories that it predicts) was known, so the Bohr model was no
longer considered that important, although it still enjoyed a certain
respect. Since the model by which Bohr achieved respect among physicists was
based on particle trajectories, it was really a surprise when Bohr in his
new interpretation proposed that particle trajectories did not exist at all.
But this was not the most radical part of his interpretation. The most
radical part was the following:
he proposed that it did not even make sense to talk
about particle properties unless these properties were measured. 
An immediate
argument against such a proposal was the well-established classical
mechanics, in which particle properties clearly existed even without
measurements. Bohr argued that there was a separation between the microscopic
quantum world and the macroscopic classical world, so that the
measurement-independent properties made sense only in the latter. However,
Bohr never explained how and where this separation took place. In his
interpretation, he introduced no new equation. His arguments were considered
pure philosophy, not physics. 
Although his arguments were partially inspired by the 
widely accepted Heisenberg
uncertainty relations, the orthodox interpretation of the uncertainty
relations (expressing practical limitations on experiments, rather
than properties of nature itself) seemed more viable. 
Thus, it is not a surprise that his
interpretation has never been taken seriously. His interpretation was soon
forgotten. (Much later it was found that the mechanism of decoherence
through the interaction with the environment provides a sort of dynamical
separation between ``classical" and ``quantum" worlds, but this separation
was not exactly what Bohr suggested.\cite{foot6})

Another prominent physicist who criticized the orthodox interpretation of QM
was Einstein. He liked the determinism of orthodox QM
(despite the fact he made contributions to the probabilistic
descriptions of quantum processes such as spontaneous
emission and photoelectric effect),
but there was
something else that was bothering him. 
To see what, consider a system containing $n$ particles 
with positions ${\bf x}_1,\ldots,{\bf x}_n$ described by a
single wave function $\psi({\bf x}_1,\ldots,{\bf x}_n,t)$.
The $n$-particle analog of (\ref{Q}) is a nonlocal function
of the form $Q({\bf x}_1,\ldots,{\bf x}_n,t)$.
In general it is a truly nonlocal function, i.e., not of the form
$Q_1({\bf x}_1,t)+\ldots +Q_n({\bf x}_n,t)$, provided
that the system exhibits entanglement, i.e., 
that the wave function is not of the form
$\psi_1({\bf x}_1,t) \cdots \psi_n({\bf x}_n,t)$.
In the orthodox interpretation such nonlocal
$Q$ is interpreted as a nonlocal
potential that determines forces on particles that depend on
{\em instantaneous} positions of all other particles.
This means that entangled spatially separated particles
must communicate instantaneously. 
Einstein argued that this is in contradiction with his theory of
relativity, because he derived that no signal can exceed the velocity of
light. Orthodox quantum physicists admitted that this is a problem for their
interpretation, but soon they found a solution. They observed that the
geometric formulation of relativity does not really exclude superluminal
velocities, unless some additional properties of matter are assumed. Thus,
they introduced the notion of {\em tachyons},\cite{foot7}
hypothetical particles that
can move faster than light and still obey the geometrical principles of
relativity. Einstein admitted that tachyons are consistent with relativity,
but he objected that this is not sufficient to solve the problem of
instantaneous communication. If the communication is instantaneous, then it
can be so only in one reference frame. This means that there must be a
preferred reference frame with respect to which the communication is
instantaneous, which again contradicts the principle of relativity according
to which all reference frames should enjoy the same rights. At that time
orthodox quantum physicists understood relativity sufficiently well to
appreciate that Einstein was right. On the other hand, the theory of
relativity was also sufficiently young at that time, so that it did not
seem too heretic to modify or reinterpret the theory of relativity itself.
It was observed that with a preferred foliation of spacetime specified by a
fixed timelike vector $n^{\mu}$ one can still write all quantum equations in
a relativistic covariant form. It was also observed that, by using an
analogy with nonrelativistic fluids, relativity may correspond only to a 
low-energy approximation of a theory with a fundamental preferred
time.\cite{foot8}
Thus, it was clear that the
preferred foliation of spacetime does not necessarily contradict the theory
of relativity
(both special and general), 
provided that the theory of relativity is viewed as an
effective theory. 
At the beginning, Einstein was not very happy with the idea that
his theory of relativity might not be as fundamental as he thought.
Nevertheless, he finally accepted that QM is irreducibly nonlocal
when he was confronted with the rigorous mathematical proof that, in QM, 
the assumption of reality existing even without measurements
is not compatible with locality.\cite{bell}      

A new crisis for orthodox QM arose with the development of 
quantum field theory (QFT). At the classical level, fields are
objects very different from particles. As QFT seemed to be a theory
more fundamental than particle QM, it seemed natural to replace the
quantum particle trajectories with the quantum field trajectories
(or more precisely, time-dependent field configurations).
However, there were two problems with this. First, from the trajectories
of fields, it was not possible to reproduce the trajectories of
particles. Second, the idea of field trajectories did not seem to work
for fermionic (anticommuting) fields. Still, the agreement with experiments
was not ruined, as all measurable predictions of QFT were actually predictions
on the properties of particles. Therefore, it seemed natural to interpret QFT
not as a theory of new more fundamental objects (the fields), but 
rather as a more accurate effective theory of particles, in which fields
play only an auxiliary role. Indeed, the divergences typical of QFT
reinforced the view that QFT cannot be the final theory, but only an effective
one.

As quantum physics made further progress, it became clear that many theories
that were considered fundamental at the beginning turned out to be merely
effective theories. 
This reinforced the dominating paradigm according to
which relativity is also an effective, approximate theory. Nevertheless,
some relativists still believed that the principle of relativity was a
fundamental principle. Consequently, they were not satisfied with the
orthodox interpretation of QM that requires a preferred foliation of
spacetime. Instead they were trying to interpret QM in a completely local
and relativistic manner. To do that, they were forced to introduce some
rather radical views of nature. In one way or another, they were forced to
assume that a single objective reality did not exist.\cite{foot9}
However, such radical interpretations were not very appreciated by
the mainstream physicists. It did not seem reasonable to crucify one of the
cornerstones not only of physics but of the whole of science (the existence of
objective reality) just to save one relatively new theoretical principle
(the principle of locality and relativity) for which there existed good
evidence that it could be only an approximate principle.\cite{foot10}
Therefore, the deterministic interpretation of QM
survived as the dominating paradigm, while the probabilistic rules of QM,
used widely in practical phenomenological calculations, were considered
emergent, not fundamental. In fact, it has been found that, 
in some cases, the
probabilistic rules cannot be derived in a simple way, so that one is forced
to use the fundamental fully deterministic theory 
explicitly.\cite{foot11}

\section{Conclusion}

In this paper, I have argued that, in the context of scientific paradigms
that were widely accepted when the Schr\"odinger equation was
discovered, it was much more natural to propose and accept the Bohmian
deterministic interpretation than the Copenhagen interpretation. I have also
argued that, if that had really happened, then the Bohmian interpretation 
(or a minor modification of it) would have been a dominating view even today. 
In other words, the answer to the allegoric tongue-twisting question posed
in the title of this paper is -- probably no!
This, of
course, does not prove that the Bohmian interpretation is more likely
to be correct than some other interpretation. But the point is that it
really seems surprising that the history of QM chose a path in which the
Copenhagen interpretation became much more accepted than the Bohmian one. I
leave it to the sociologists and historians of science to explain why the
history of QM chose the path that it did.  

\section*{Acknowledgments}
The author is grateful to anonymous referees for numerous suggestions
for improvements.
This work was supported by the Ministry of Science of the
Republic of Croatia under Contract No.~098-0982930-2864.


\begin{thebibliography}{99}

\bibitem{foot0}
A similar thesis with somewhat 
different arguments has been also advocated in 
J. T. Cushing, 
Quantum Mechanics: Historical Contingency and the Copenhagen Hegemony
(University of Chicago Press, Chicago, 1994).

\bibitem{foot1}
Remarks concerning the actual history of QM
are given in references.

\bibitem{bohm}
D. Bohm,
``A suggested interpretation of the quantum theory in terms of
``hidden variables". I,"
Phys. Rev. {\bf 85} (2), 166-179 (1952);
D. Bohm,
``A suggested interpretation of the quantum theory in terms of
``hidden variables". II,"
Phys. Rev. {\bf 85} (2), 180-193 (1952).

\bibitem{tumul}
R. Tumulka,   
``Understanding Bohmian mechanics: A dialogue,"
Am. J. Phys. {\bf 72} (9), 1220-1226 (2004).

\bibitem{foot2}
Such an interpretation was really proposed already in 1926:
E. Madelung, 
Z. Phys. {\bf 40}, 322-326 (1926).

\bibitem{foot3}
These arguments might have looked similar to those in
D. D\"urr, S. Goldstein, and N. Zangh\`i,
``Quantum equilibrium and the origin of absolute uncertainty,"
J. Stat. Phys. {\bf 67}, 843-907 (1992);
A. Valentini,
``Signal-locality, uncertainty, and the subquantum H-theorem,"
Phys. Lett. A {\bf 156}, 5-11 (1991). 

\bibitem{foot4}
In reality, this interpretation is known today as
the Bohm interpretation, while the status of an ``orthodox" interpretation
is enjoyed by a significantly different interpretation.
De Broglie has also proposed the same equation
for particle trajectories 
much earlier than Bohm did, 
but de Broglie did not develop
a theory of quantum measurements, so he could not reproduce the
predictions of standard QM for observables other than particle
positions, such as particle momenta. For more historical details
see also
G. Bacciagaluppi and A. Valentini, 
Quantum Theory at the Crossroads: Reconsidering the 1927 Solvay Conference
(to be published by Cambridge University Press);
quant-ph/0609184.



\bibitem{foot5}
Such arguments might have looked similar to
those in
H. Nikoli\'c, 
``Classical mechanics without determinism,"
Found. Phys. Lett. {\bf 19}, 553-566 (2006).
%
In this paper, it is shown that classical statistical physics can be
represented by a nonlinear modification of the Schr\"odinger
equation, in which classical particle trajectories may be identified
with special solitonic solutions. A Bohr-like interpretation
of general (not solitonic) solutions suggests that even classical
particles may not have trajectories when they are not measured,
while a measurement of the previously unknown
position may induce an indeterministic wave-function 
collapse to a solitonic state.

\bibitem{foot6}
For a review of the theory of
decoherence with 
emphasis on the interpretational issues, see
M. Schlosshauer,
``Decoherence, the measurement problem, and interpretations of quantum
mechanics,"
Rev. Mod. Phys. {\bf 76}, 1267-1305 (2004).

\bibitem{foot7}
In reality, tachyons
have been introduced in physics somewhat later, see
O. M. P. Bilaniuk, V. K. Deshpande, and E. C. G. Sudarshan,
````Meta" relativity,"
Am. J. Phys. {\bf 30} (10), 718-723 (1962);
%
O. M. P. Bilaniuk and E. C. G. Sudarshan,
``Particles beyond the light barrier,"
Physics Today {\bf 22} (5), 43-51 (1969).

\bibitem{foot8}
%
It is well known that a wave equation describing the propagation of sound
with the velocity $c_s$ in a fluid has the same mathematical form as a 
special-relativistic
wave equation describing the propagation of light with the velocity $c$
in vacuum.
Consequently, such a wave equation of sound is invariant with respect to Lorentz
transformations in which the velocity $c$ is replaced by $c_s$.
A fluid analogy of curved spacetime may also be constructed, by 
introducing an inhomogeneous fluid. 
%
For more details, see, e.g., 
M. Visser,
``Acoustic black holes: horizons, ergospheres, and Hawking radiation,"
Class. Quant. Grav. {\bf 15}, 1767-1791 (1998).



\bibitem{bell}
This proof is now usually attributed to Bell, although
other versions of this proof also exist. For a 
pedagogic review see 
F. Lalo\"e, ``Do we really understand quantum mechanics?
Strange correlations, paradoxes, and theorems," 
Am. J. Phys. {\bf 69} (6), 655-701 (2001).

\bibitem{foot9}
Many of the
current interpretations of QM mentioned in the introduction are of this
form.

\bibitem{foot10}
String
theory also contains evidence against locality at the fundamental level.
%
Although the theory is originally formulated as a local theory, nonlocal
features arise in a rather surprising and counterintuitive manner.
It turns out that string theories defined on different background spacetimes
may be mathematically equivalent, which suggests that spacetime is not 
fundamental at all. Without a fundamental notion of spacetime,
there is no fundamental notion of locality and relativity as well. 
It is believed that
a more fundamental formulation of string theory should remove locality
more explicitly, while known local laws of field theory should emerge
as an approximation.
%
See, e.g.,
G. T. Horowitz,
``Spacetime in String Theory,"
New J. Phys. {\bf 7}, 201 (2005);
N. Seiberg,
``Emergent Spacetime,"
hep-th/0601234.

\bibitem{foot11}
%
It is known that relativistic QM based on the Klein-Gordon equation,
as well as QFT, do not contain a position operator. Therefore, the conventional
interpretation of quantum theory does not have clear predictions on
probabilities of particle positions in the relativistic regime.
The fundamentally deterministic Bohmian interpretation 
may lead to clearer predictions, which means
that it may be empirically richer than (and thus inequivalent to) the conventional formulation. 
%
For more details, see, e.g., 
H. Nikoli\'c, 
``Relativistic quantum mechanics and the Bohmian interpretation,"
Found. Phys. Lett. {\bf 18}, 549-561 (2005);
H. Nikoli\'c, ``Is quantum field theory a genuine quantum theory?
Foundational insights on particles and strings," arXiv:0705.3542.
%
Unfortunately, experiments that could confirm or reject such a formulation 
have not yet been performed.
It is also fair to note that today such a version of the Bohmian interpretation
not empirically equivalent to the conventional interpretation is considered controversial
even among the proponents of the Bohmian interpretation. Nevertheless, 
in an alternative history of QM in which the conventional probabilistic
interpretation never became widely accepted, such a fundamentally 
deterministic Bohmian interpretation might have seemed more natural.


\end{thebibliography}
\end{document}